\documentclass[
showpacs,
floatfix,
aps,
pra,
amsmath,
twocolumn,
groupaddress,
]{revtex4-1}

\usepackage{amsmath,amssymb}
\usepackage{amscd, latexsym}
\usepackage{mathrsfs}
\usepackage{graphicx}
\usepackage{epstopdf}
\usepackage{cancel}
\usepackage{amsfonts}
\usepackage{exscale}
\usepackage{dcolumn}
\usepackage{bm}
\usepackage{color}
 \usepackage{natbib}
\usepackage{lmodern}
\usepackage[T1]{fontenc}
\usepackage[latin9]{inputenc}
\usepackage{verbatim}
\usepackage{float}
\usepackage {hyperref}
\hypersetup{colorlinks=true}

\newcommand{\ket}[1]{\left| #1 \right>} 
\newcommand{\bra}[1]{\left\langle #1 \right|}

\begin{document}

\title{Response to a local quench of a system near many body localization transition}

\author{Canran Xu and Maxim G. Vavilov}

\affiliation{Department of Physics, University of Wisconsin-Madison, Wisconsin 53706, USA }

\date{\today}

\begin{abstract} 

We consider a one dimensional spin $1/2$  chain with Heisenberg interaction  in a disordered parallel magnetic field.  This system is known to exhibit the many body localization (MBL) transition at critical strength of disorder.  We analyze the response of the chain when additional perpendicular magnetic field  is applied to an individual spin and propose a method for accurate determination of the mobility edge via local spin measurements.  We further demonstrate that the exponential decrease of the spin response with the distance between perturbed spin and measured spin can be used to characterize the localization length in the MBL phase. 

\end{abstract}

\maketitle 
{\em Introduction.--}\label{sec1}
Theoretical studies of localization in many body systems were originally focused on interacting electrons in disordered metals \cite{Basko2006,Gornyi2005,Oganesyan2007}, 
where
experimental observation of localization remains a challenging task due to electron-phonon interaction that ruins many-particle states by introducing external decoherence to the system of interacting electrons in real metals.  The interest in observation of many-body localization (MBL) has shifted to artificial quantum systems well isolated from their environment  such as ultra-cold  atomic gases \cite{Schreiber2015}, trapped ions \cite{Senko2014, Islam2013} and superconducting circuits containing many interacting qubits \cite{Mutus}.  These systems have configurable Hamiltonians and their observables can be measured with high precision, which gives another advantage to artificial quantum systems over their solid state counterparts for studies of MBL.

Artificial quantum systems containing relatively small number of quantum particles are also attractive from theory side as they can be simulated numerically by exact diagonalization of their Hamiltonians or by approximate methods. 
Recent theoretical work was focused on numerical and analytical studies of interacting one dimensional spins chains~\cite{Serbyn2013b, Chandran2015,Ros2015420, Kim2014,Bardarson2012, Serbyn2013, Nanduri2014, Serbyn2014, Vosk2014b, Kjall2014, DeRoeck2015} that 
has shown that spin systems containing more than ten spins and  involving thousands of many-body eigenstates exhibit the MBL behavior in sufficiently strong disorder.  
The MBL phase can be characterized by the existence of infinite number of local integrals of motion \cite{Serbyn2013b, Chandran2015,Ros2015420, Kim2014}, the entanglement~\cite{Bardarson2012, Serbyn2013, Nanduri2014, Serbyn2014, Vosk2014b, Kjall2014}, as well as the spectral properties of eigenstates~\cite{Oganesyan2007}.  While these characteristics can be measured in principle, the corresponding experiments are strenuous  as they require a full quantum tomography of systems' states.

In this paper we propose an alternative strategy to identify the localization in a disordered system of interacting spins. We analyze the response of 
a pure state 
of a Heisenberg spin chain in random magnetic field along $z$ axis to a sudden application of a magnetic field (quench) perpendicular to $z$ that acts on a single spin.  We evaluate the inverse participation ratio (IPR) of an eigenstate of unperturbed Hamiltonian in the basis of the quenched Hamiltonian.
The IPR is small in the delocalized, or ergodic, regime when the initial state overlaps with many eigenstates of the new Hamiltonian.  In the  localized regime, an application of a quench does not affect majority of eigenstates and the typical value of the IPR is about unity.   Points at which the IPR starts increasing fast form a curve in the energy \textit{vs} disorder strength plane and define 
the mobility edges. 
We note that due to finite size of our system, the mobility edge cannot be defined rigorously and the mobility edge has to be treated as a crossover region.  Recently, the authors of Ref.~\cite{DeRoeck2015} argued that even in infinite system MBL and ergodic phases are separated by a crossover region.

Since the IPR is not easily measurable in experiments, we also investigate correlations in single spin measurements before and after the quench.  The covariance between these two measurements is small for delocalized states, but rapidly increases for localized states, as the quench only weakly affects configuration of far away  spins.  The mobility edge obtained from the covariance is consistent with the mobility edge obtained through the IPR, as well as through analysis of the entanglement entropy \cite{Luitz2015, Shem2015}.  Moreover, at strong disorder in the localization regime, the spin response to the quench decreases exponentially as a function of the distance of the monitored spin from the quenched spin. 
We utilize this exponential decay to evaluate the localization length as a function of disorder and demonstrate that the localization length exceeds the system size at the onset of MBL.

\begin{figure}[h]
\begin{centering}
\includegraphics[width=7.5cm]{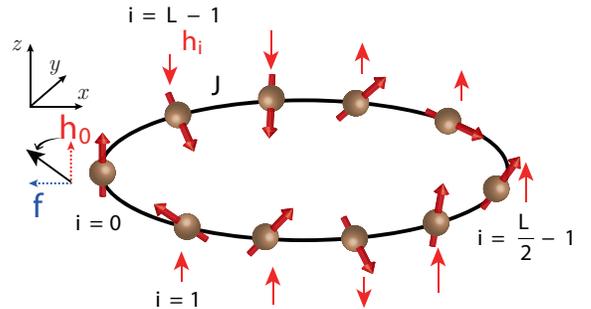}
\par\end{centering}
\caption{\label{fig:illu} 
(Color online) Schematic picture of the one-di\-men\-sional spin-1/2 chain with periodic boundary condition. Along the chain, each spin is subject to a random onsite field $h_i$ along $z$ direction and the spins are coupled by nearest neighbor Heisenberg interactions with strength $J$. At time $t_0$ when the local quench is turned on, a transverse magnetic field $f$ is applied to one of the spin, labeled by $i=0$.
}

\end{figure}

{\em Local Quench.--}\label{sec2}
We consider a Heisenberg chain of $L$ spins with random on-site field in the $z$ direction and  periodic boundary conditions:
\begin{equation}
H_0 = \sum_{i=0}^{L-1} h_i\sigma_z^{(i)} + J\sum_{i=0}^{L-1}  \bm{\sigma}^{(i)}\cdot\bm{\sigma}^{(i+1)},\label{eq:H0}
\end{equation}
where $h_i$ on each site is a random variable distributed uniformly in the interval $[-W, W]$ and $\bm{\sigma}^{(i)}$ is the Pauli matrix for spin at site $i$, see Fig.~1. Throughout the paper, we use $J$ as a fundamental unit and set $J=1$.  We denote eigenstates of $H_0$ by $\ket{\alpha_{S_z}}$, where the total spin in $z$ direction $S_z = \sum_i \sigma_z^{(i)}$ is conserved for Hamiltonian~\eqref{eq:H0}.

Previous numerical \cite{Pal2010, Kjall2014} and analytical \cite{Vosk2014} studies were focused on the subspace defined by energy states 
$\ket{\alpha_{0}}$ with $S_z=0$, where 
the MBL phase develops at $W\gtrsim 3.4$.
In this paper, we study the effect of the sudden quench $V = f\sigma_x^{(i=0)}$, see Fig.~\ref{fig:illu}, applied to Hamiltonian \eqref{eq:H0} at site $i=0$. The new Hamiltonian $\tilde H = H_0 + V$ break conservation of $S_z$, and system dynamics occur in the  full $2^L$ dimensional space with the basis 
defined by  eigenvectors $\ket{\tilde{\alpha}}$ of $\tilde H$. The response of the system to this quench is reminiscent to the quasiparticle spectral function that appears in problems about transport in disordered electron systems with interactions.~\cite{Basko2006a}

We consider a system that  was originally prepared as a pure state $\ket{\psi}$ in the subspace of states with $S_z \ket{\psi} =0$ and calculate its response over a long time after the quench:
\begin{equation}
\label{eq:O}
\bar{\langle O\rangle} = \lim_{T\rightarrow\infty} \frac{1}{T} \int_0^T \hat O(t) dt = {\rm Tr}\overline{\rho(t) \hat O} = {\rm Tr}{\rho_{\rm DE}^{\ket{\psi}} \hat O},
\end{equation}
where $\rho(t) = \exp{(-i\tilde Ht)}\rho(0)\exp{(i\tilde Ht)}$ is the density matrix 
and $\rho_{\rm DE}^{\ket{\psi}}= \overline{\rho(t)}$ is the time-averaged  density matrix initialized in a pure state.  Then,  
$\rho(0)=\ket{\psi}\bra{\psi}$,
\begin{equation}
\rho_{\rm DE}^{\ket{\psi}} = \sum_{\tilde\alpha}^{2^L} P_{\tilde\alpha}   \left|
\bra{\tilde\alpha}\psi\rangle\right|^2, \quad P_{\tilde\alpha} = \ket{\tilde\alpha}\bra{\tilde\alpha}
\end{equation} 
and $P_{\tilde\alpha}$
is the projection operator on new eigenstates $\ket{\tilde\alpha}$ of 
$\tilde H$. The off-diagonal elements of $\rho_{\rm DE}^{\ket{\psi}}$ vanish after time  averaging and matrices  $\rho_{\rm DE}^{\ket{\psi}}$  belong to a diagonal 
ensemble~\cite{Rigol2008}.

Below we concentrate on initial states $\ket{\psi}$ that are eigenstates $\ket{\alpha_0}$ of the initial Hamiltonian $H_0$ with $S_z=0$, here level index $\alpha_0=1,\dots N(L,0)$ runs over eigenenergies of $H_0$ ordered in increasing order, $N(L,S)=(2^L)!/[(2^{(L-S)/2})!(2^{(L+S)/2})!]$.

{\em  Inverse participation ratio --}\label{sec3}
We 
demonstrate that the Hamiltonian $\tilde H$ with the quench exhibits the MBL phase by analyzing the IPR:
\begin{equation}
{\rm IPR}_{\alpha_0} = \sum_{\tilde{\alpha}=1}^{2^L}|\langle \alpha_0|\tilde{\alpha}\rangle|^4 = {\rm Tr}\left\{\bar{\rho}^{\ket{\alpha_0}}_{\rm DE} P_{\alpha_0}\right\},
\end{equation}
where $P_{\alpha_0} = \ket{\alpha_0}\bra{\alpha_0}$.  The IPR is a measure of portion of the Hilbert space explored by the system after the perturbation $V$ is turned on  and was introduced earlier for analysis of mobility edge in disordered electron systems with interactions~\cite{Basko2006a,deLuca2013}. 
At weak disorder, the motion of the system is ergodic and the state spreads over a large fraction of the Hilbert space.  As a result, the IPR is small $\sim 2^{-L}$.  On the other hand, in the strong disorder limit, the ergodicity breaks down and the evolution of many-body wavefunctions is restricted to a small portion of the Hilbert space defined by the local integral of motion~\cite{Serbyn2013b}.

\begin{figure}[h]
\begin{centering}
\includegraphics[width=9cm]{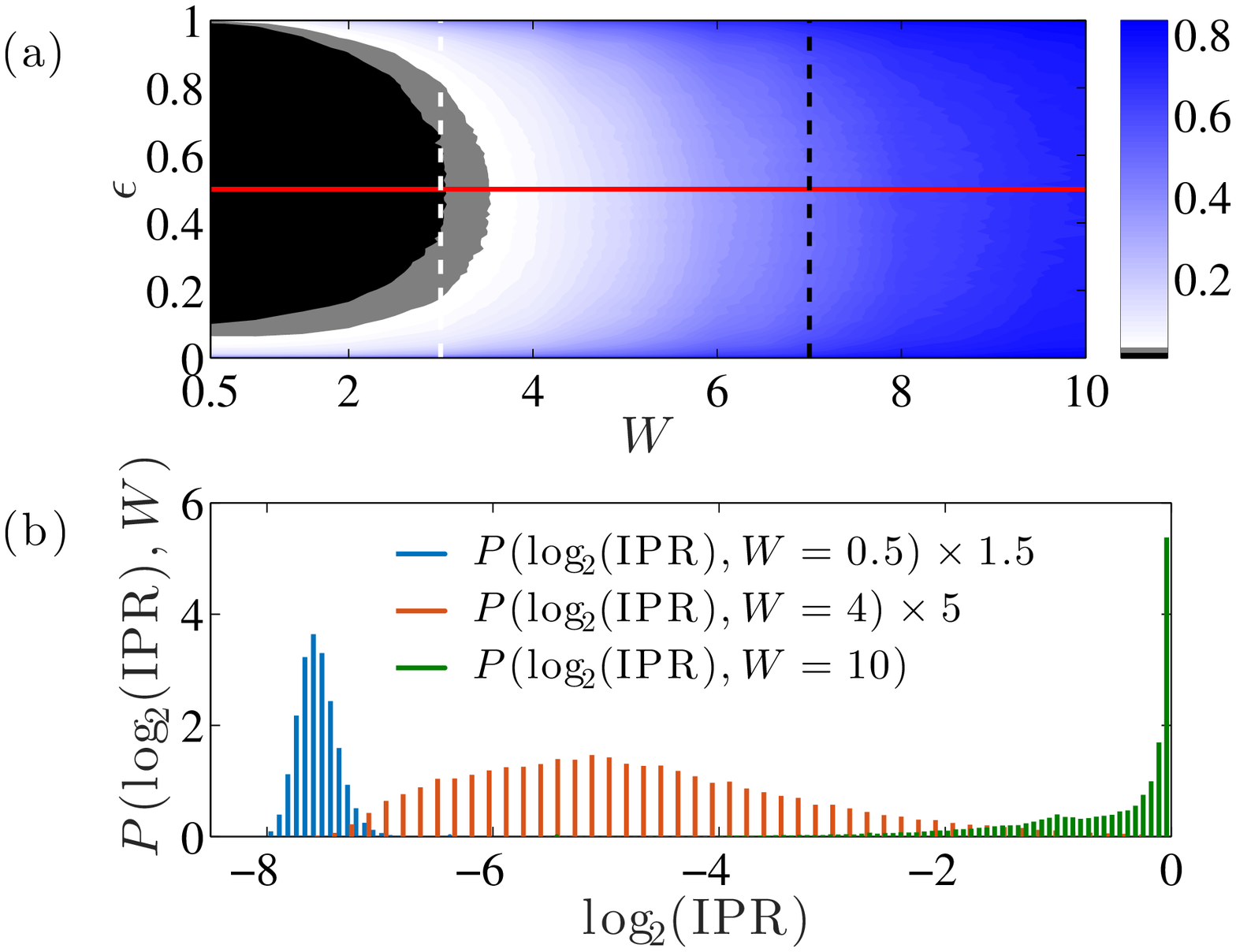}
\includegraphics[width=9cm]{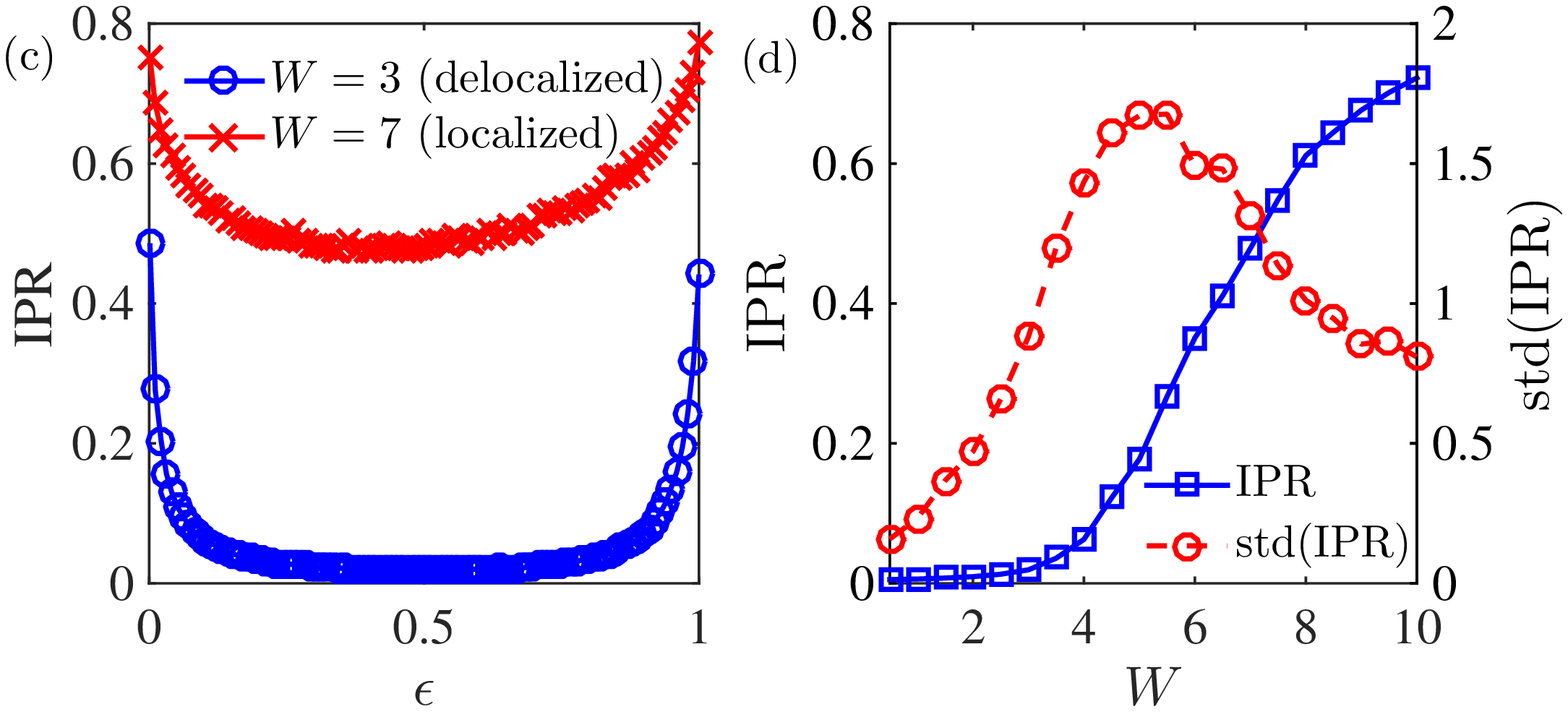}
\par\end{centering}
\caption{\label{fig:P_r}
(Color online) (a) Average IPR over 1000 disorder realizations as functions of disorder strength $W$ and energy density $\epsilon$ for a system of $L = 12$  and $f = J=1$. The many body mobility edge along the outer border of the light-gray sector encloses a region of ergodic states with ${\rm IPR}\sim 2^{-L}$. The horizontal and vertical lines indicate parameters for panels (c) and (d). (b) The histogram of $\log_2{\rm IPR}$ in the middle of the band for $W = 0.5$, $4$ and $10$. The distributions are narrow  for weak and strong disorder, but broad near the critical region. (c) The ${\rm IPR}$ at fixed disorder $W = 3$, when the states in the middle of the band are ergodic but localized near the band edges, and $W = 7$ when all states are localized. (d) The ${\rm IPR}$ and its standard deviation for fixed energy density $\epsilon \simeq 0.5$.
}
\end{figure}

To investigate behavior of the IPR across the MBL transition we performed exact diagonalization for  $L = 12$ spins and  $N = 1000$ realizations to obtain all eigenstates $\ket{\alpha_0}$ for $H_0$ and $\ket{\tilde{\alpha}}$ for $\tilde H$.  The quench strength is fixed at $f = J$ throughout our computations.   
The average ${{\rm IPR}}$  as a function of disorder strength $W$ and place of eigenenergy within the energy band $\epsilon = \alpha_0/N(L,0)$ is plotted in Fig.~\ref{fig:P_r}(a). 
After disorder averaging ${{\rm IPR}}(W, \epsilon)$  clearly reveals the existence of a mobility edge that distinguishes ergodic from localized states. To justify the nature of the mobility edge,  we plot the histogram of the distribution of $\log_2{\rm IPR}$ in Fig.~\ref{fig:P_r}(b). In the weak and strong disorder limit, the distributions of $\log_2{\rm IPR}$ are highly concentrated at either $\propto (-L)$ or 0, respectively. However, $\log_2{\rm IPR}$ in the crossover region is broadly distributed between $(-L)$ and 0 with its standard deviation $\propto L$ so that the standard deviation is peaked at the mobility edge. 
This singular behavior is similar to that of the fluctuations of the entanglement entropy~\cite{Kjall2014} and  the form of the IPR distribution at weak and strong disorder resembles the distribution of the quasiparticle spectral functions~\cite{Basko2006, Basko2006a}.  

In the lower panels Fig.~\ref{fig:P_r}(c), we make two vertical cuts at fixed disorder strengths $W=3$, 7 on the phase diagram.
For moderate disorder strength $W = 3$ and in the presence of mobility edge, the ${\rm IPR}$ reaches $\sim 0.4$ at the edges in an unsymmetrical manner, but it sharply drops to $\sim 10^{-2}$ and forms a flat basin of ergodic states in the middle of the band. At strong disorder $W = 7$, ${\rm IPR}_{\alpha_0}\gtrsim 0.4$, indicating that most states are localized.
When we fix the energy of an initial state in the middle of the band, $\epsilon \simeq 0.5$, the standard deviation of IPR has a peak at the critical value $W_c \simeq 3.5$, when the upper and lower mobility edges merge, see Fig.~\ref{fig:P_r}(d).  
We note that at strong disorder, $W\simeq 10$, the standard deviation of IPR remain of the order of unity, see Fig.~\ref{fig:P_r}(b,d).  indicating a broad distribution of the IPR in the localized regime.  This behavior of the IPR in the middle of the energy band can be connected to the statistical orthogonality catastrophe~\cite{Deng2015,Khemani2015}.

{\em Local observables --}\label{sec4}
While the above approach to detect mobility edge through IPR is suitable for numerical calculations, it is hardly realized experimentally.
For experimental detection of the mobility edge, we propose a different approach.  We study the dynamics of single spins and investigate the correlation of their expectation values $\langle {\sigma}_z^{(i)}\rangle$ before and after a local perturbation. 
Our motivation is based on the previous observations\cite{Nandkishore2015} that in the MBL regime the eigenstate thermalization hypothesis (ETH) is violated and information about the local integrals of motion maintains correlations between spin states before and after the quench.  Otherwise, at weak disorder the motion is ergodic and  information about initial states is lost.

\begin{figure}[t]
\begin{centering}
\includegraphics[width=9cm]{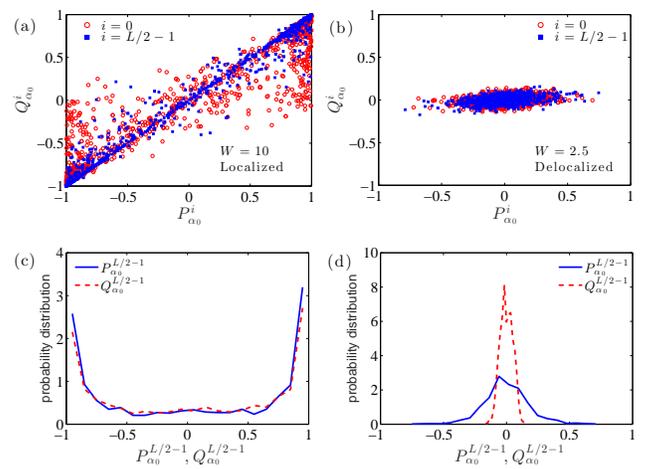}
\par\end{centering}
\caption{\label{fig:z1z2}(Color online) Scatter plot of ($P_{\ket{\alpha}}^i$, $Q_{\ket{\alpha}}^i$) for (a) the localized regime with $W = 10$ and (b) the ergodic regime with $W = 2.5$ for states $\ket{\alpha_0}$  in the middle of the band. Data obtained with $L = 12$ and $N = 1000$. The corresponding probability distributions of $P_i$ and $Q_i$ for the localized regime (c) in which the single spin measurement with and without the quench are similar and for the  ergodic regime (d) the discrepancy in the distribution with and without the quench indicates thermalization.}
\end{figure}

For an interacting system, the observables set by a finite degrees of freedom can be evaluated by the reduced density matrices in which the off-diagonal elements are essentially zero due to dephasing even if the system starts in some arbitrary pure state.  Alternatively, time averaged expectation values of a local operator are also characterized by a diagonal ensemble of the density matrix $\rho_{\rm DE}^{\ket{\psi}}$, see Eq.~\eqref{eq:O}. 
One can measure such local observables before and after the quench.  Below, we again consider a special case of initial states that are eigenstates $\ket{\alpha_0}$ of $H_0$ with $S_z=0$, although the results remain qualitatively the same for a system that is initially prepared in an arbitrary pure state.
We calculate expectation values of $\langle \sigma_z^{(i)}\rangle$ of spin $i$ 
before and after the quench at site $0$:
\begin{subequations} 
\begin{align}
P_{\alpha_0}^i =& \bra{\alpha_0}\sigma_z^{(i)}\ket{\alpha_0} 
,\\
Q_{\alpha_0}^i =& \sum_{\tilde{\alpha}}|\langle \alpha_0 | \tilde{\alpha}\rangle|^2\bra{\tilde{\alpha}}\sigma_z^{(i)}\ket{\tilde{\alpha}} = {\rm Tr}
\left\{\bar{\rho}_{\rm DE}^{\ket{\alpha_0}}\sigma_z^{(i)}
\right\},
\end{align}
\end{subequations}
where 
$\rho_{\rm DE}^{\ket{\alpha_0}}= \sum_{\tilde\alpha}P_{\tilde{\alpha}} \left|\langle \tilde\alpha|\alpha_0\rangle\right|^2$.

A good distinction between ergodic and MBL phases can be obtained by analyzing correlations between measurements of $P^i_{\alpha_0}$ and $Q_{\alpha_0}^i$. 
Running over all eigenstates $\ket{\alpha_0}$ in the $S_z = 0$ sector, we collect $P_{\alpha_0}^i$ and $Q_{\alpha_0}^i$ 
for a number of disorder realizations and present the scatter plot for pairs of ($P_{\alpha_0}^i$, $Q_{\alpha_0}^i$) of spin $i=0$ (directly perturbed spin, open circles) and $i=L/2-1$ (the farthest spin from the quench, filled squares) in the middle of the band, $\alpha_0\simeq N(L,0)/2$ for strong ($W = 10$) and weak ($W = 2.5$) disorder, see Fig.~\ref{fig:z1z2}. 
In the localized phase at strong disorder, the eigenstates are product states consisting of physical spins $\ket{\alpha} = \bigotimes_i\ket{\downarrow(\uparrow)}_i$ and therefore the local spin projection is good quantum number $P_{\alpha_0}^i \simeq \pm 1$. 
Provided that the quench is smaller than  the local random field, the eigenstates of $H_0$ and $\tilde H$ differ by terms $\mathcal{O}(f/W)$, resulting in an almost unchanged  $Q_{\alpha_0}^i \simeq P_{\alpha_0}^i$, when the distributions of $P_{\alpha_0}^i$ and $Q_{\alpha_0}^i$ are peaked at $\pm 1$, see Fig.~\ref{fig:z1z2}(c). 
On the scatter plot, pairs  
($P_{\alpha_0}^{L/2-1}$, $Q_{\alpha_0}^{L/2-1}$) are distributed along the line $P_{\alpha_0}^{L/2-1}=Q_{\alpha_0}^{L/2-1}$, indicating the two measurements are strongly correlated. 
In the ergodic phase at weak disorder ($W = 2.5$), the eigenstate is a linear combination of a large number of product states, and therefore local spin projection is not a good quantum number. Two sets of local spin  measurements form an elliptic cloud, see Fig.~\ref{fig:z1z2}(b). This distribution points to ergodicity: the distribution of single spin measurement for all possible spin configurations have maxima at $P_{\alpha_0}^i = 0$ and $Q_{\alpha_0}^i = 0$ and the distribution of $Q_{\alpha_0}^i$ is narrower due to the equilibration of the system after the quench, see Fig.~\ref{fig:z1z2}(d) for $W = 2.5$.
Remarkably, the distributions are almost indistinguishable for MBL regime at $W = 10$, \textit{cf.} Figs.~\ref{fig:z1z2}(c,d).

\begin{figure}[h]
\begin{centering}
\includegraphics[width=8.5cm]{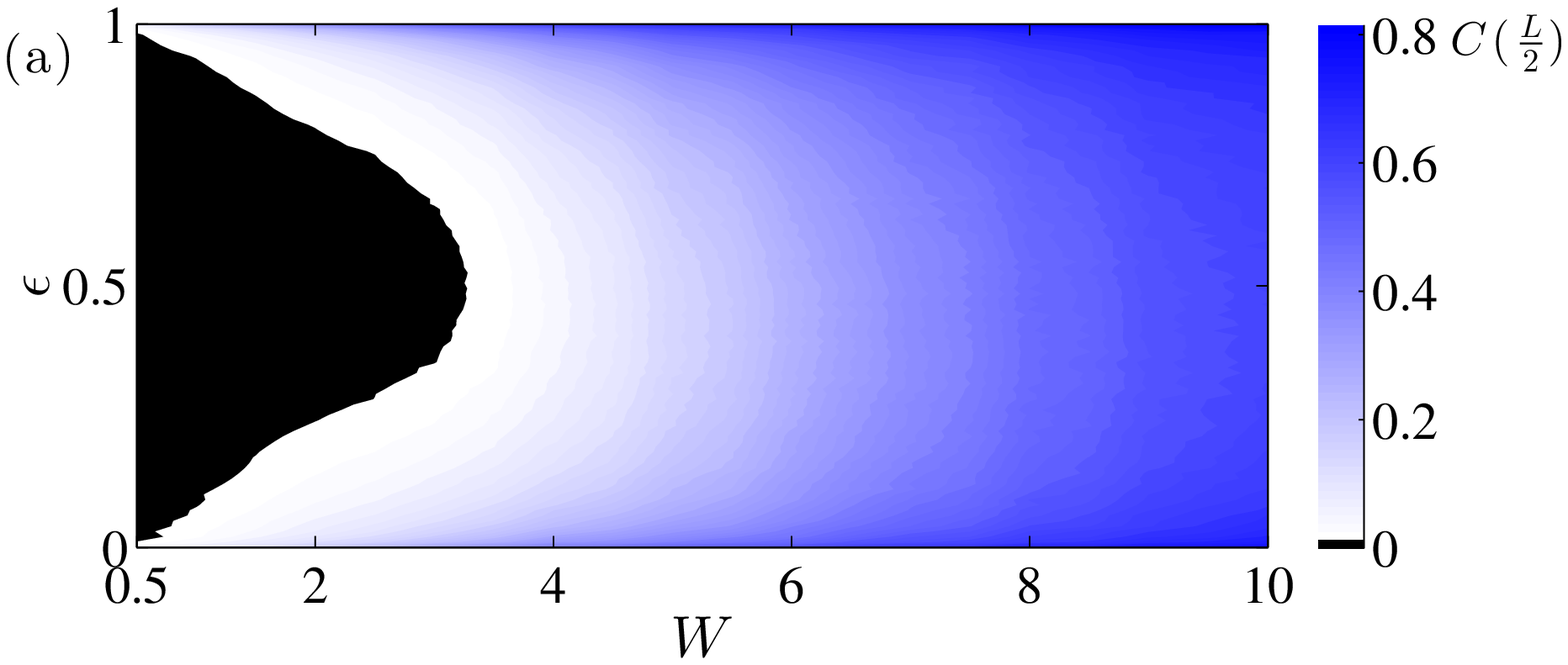}
\includegraphics[width=8.5cm]{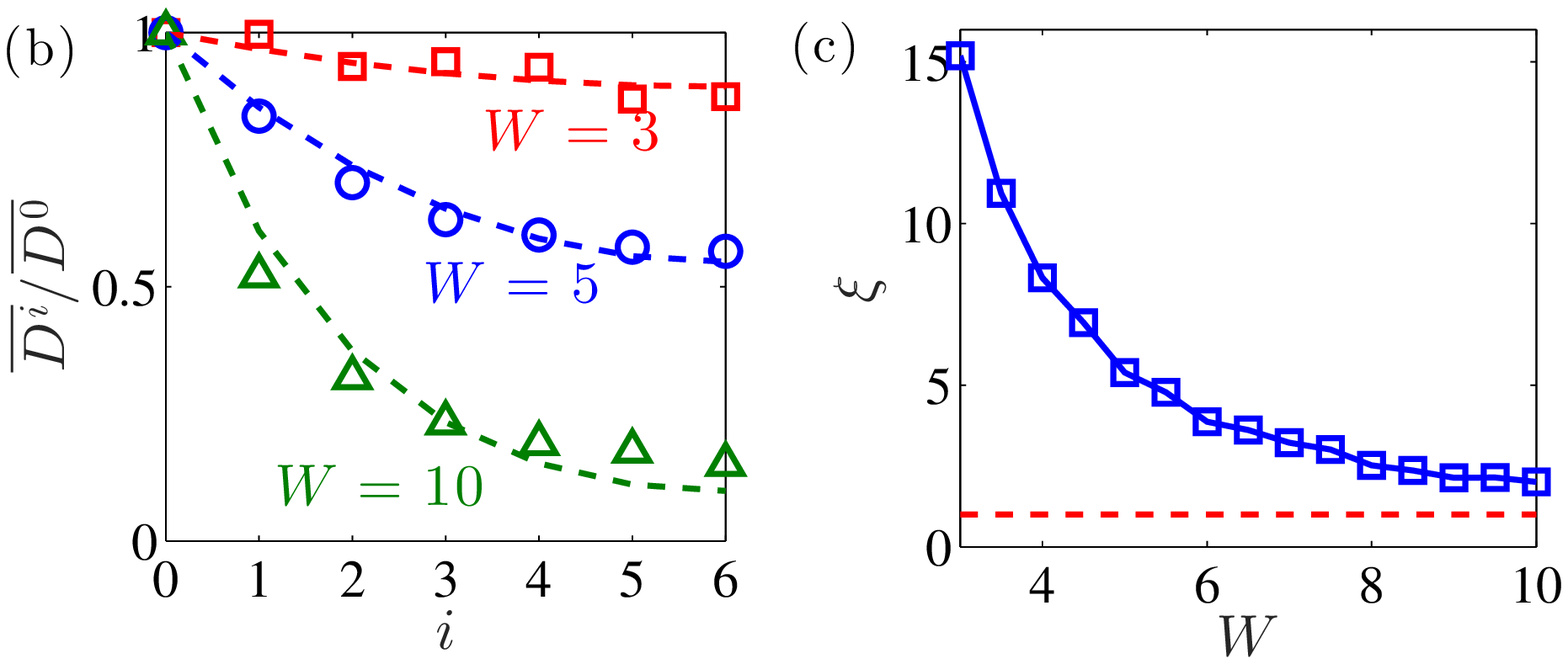}
\par\end{centering}
\caption{\label{fig:z1z2xi}(Color online) (a) Correlator $\overline{C_{\ket{\alpha}}^i} = \overline{P_{\ket{\alpha}}^iQ_{\ket{\alpha}}^i}$ as a function of disorder $W$ and energy density $\epsilon$ for $L = 12$ and $2000$ realization. The many-body mobility edge is marked in white. (b) $D^i$, see eq.~\eqref{eq:Di}  as a function of distance between the monitored and quenched spins (points), and the corresponding fitting curve, Eq.~\eqref{eq:d_exp} (dashed lines). (c) The many-body localization length $\xi$ extracted from  $D^{(i)}_{\alpha_0}$ as a function of disorder strength in the middle of the band, $\alpha_0\simeq N(L,0)/2$. The dashed line corresponds to $\xi = 1$ to show that the at strong disorder the system is localized on the atomic scale.
}
\label{fig:4}
\end{figure}

Correlations between $P_{\alpha_0}^i$ and $Q_{\alpha_0}^i$ are characterized by the covariance with respect to disorder realizations:
${C_{\alpha_0}}(i) = \overline{(P_{\alpha_0}^i - \overline{P_{\alpha_0}^i})\cdot(Q_{\alpha_0}^i - \overline{Q_{\alpha_0}^i})}
 =  \overline{P_{\alpha_0}^iQ_{\alpha_0}^i}$,
where $\overline{P_{\alpha_0}^i}\simeq \overline{Q_{\alpha_0}^i}\simeq 0$.  We use the covariance to map out the phase diagram as a function $W$ and $\epsilon=\alpha_0/N(L,0)$, as shown in Fig.~\ref{fig:z1z2xi}. In the ergodic regime the averaged value ${C_{\alpha_0}}(L/2)$ vanishes but it saturates to 1 deep in the MBL phase where both ${P_{\alpha_0}^i}$ and ${Q_{\alpha_0}^i}$  take almost identical values $\sim (\pm 1)$.  Similar to the IPR, $C_{\alpha_0}(i)$ reveals the many-body mobility edge, marked by the border of the black region in Fig.~\ref{fig:4}(a). 

With decreasing the spatial separation between the quenched spin and the monitored spin, we observe larger deviations between $P_{\alpha_0}^i$ and 
$ Q_{{\alpha_0}}^i$ as the perturbed spin can be thermalized with subsystem of surrounding spins within a cluster of a size of the localization length. 
On the contrary, in the ergodic regime, the distribution of ($P_{{\alpha_0}}^i$, $Q_{{\alpha_0}}^i$) is insensitive to the spatial separations between quench at site 0 and monitored spin $i$, suggesting that the thermolization happens throughout the whole system. 
The ensemble statistics of $P_{{\alpha_0}}^i$ and $Q_{{\alpha_0}}^i$ can be used to evaluate the localization length. 
Intuitively the localization length is a scale at which spin texture form localized clusters and thereby the ergodicity is broken. Essentially deep in the localized regime, localization occurs on atomic scales  with  $\xi \to 1$.   With decreasing disorder, the localization length grows and once the scale is beyond the system size, the entire system cannot be decomposed into independent clusters and the ergodic behavior takes place. Therefore, the localization length is an indicator of the onset of MBL regime that can be determined by the spatial sensitivity of the response to the local quench. The deviation between the measurement $P_{{\alpha_0}}^i$ and $Q_{{\alpha_0}}^i$ averaged over disorder realizations is given by the ``distance'':
\begin{equation}
\label{eq:Di}
D^{i}_{{\alpha_0}} = \sqrt{\overline{(P_{{\alpha_0}}^i-Q_{{\alpha_0}}^i)^2}}.
\end{equation}
We argue that in the localized regime $D^{i}_{{\alpha_0}}$ is an exponentially decaying function with respect to the distance between spin $i$ to the quenched spin:
\begin{equation}\label{eq:d_exp}
{D^{i}_{{\alpha_0}}}/{D^0_{{\alpha_0}}} \simeq e^{-i/\xi} + e^{-|L -i|/\xi} - e^{-L/\xi},
\end{equation} 
where $\xi$ is the localization length. The first two exponential terms in Eq.~\eqref{eq:d_exp} arise from the periodic boundary condition, and the last term is a normalization constant.  In Fig.~\ref{fig:z1z2xi}(b) we present the ratio ${D^i}/{D^0}$ and its best fit with respect to choice of $\xi$ in Eq.~\eqref{eq:d_exp} as a function of spatial separation $i$ for several different disorder strengths $W = 3, 5, 10$ in the middle of the band. 
In Fig.~\ref{fig:z1z2xi}(c), we illustrate  the  fitted localization length $\xi$ for different values of disorder $W$. Due to the finite size of the system, at the critical disorder $W_c \simeq 3.4$ the localization length $\xi$ does not diverge but becomes larger than the system size, consistent with our argument about the crossover between ergodic and MBL phases. On the other hand, $\xi$ saturates to unity  at strong disorder deep in the MBL phase.

{\em Summary --} We showed that as a result of local quench, the absence of thermalization in the MBL phase can be characterized by the IPR of the eigenstates of the original Hamiltonian in the basis of the quenched Hamiltonian. In particular, the IPR  fluctuations are enhanced at the crossover between  the ergodic and MBL phases. Our analysis of the single spin measurements in response to a quench shows a plausible experimental technique to search for the mobility edge and the localization length.
This approach does not require substantial measurements of a quantum system as compared to more complicated measurement of many-particle entanglement. 
Our analysis demonstrated that for a system with MBL behavior, simple single spin measurement can reveal the indispensable characteristics, complementary to more sophisticated routes based on the growth of entanglement entropy\cite{Bardarson2012, Serbyn2013} or quantum revivals\cite{Vasseur2014}.

\begin{acknowledgments}
We thank D. Basko, D. Huse, L. Ioffe, R. Nandkishore, V. Oganesyan and P. Woelfle  for fruitful discussions. This work was supported by NSF Grant No. DMR-0955500.
\end{acknowledgments}

\end{document}